\documentclass[10pt,twocolumn,prl,aps,amssymb,amsmath,superscriptaddress]{revtex4}
\include{gyrokinetics-macros}
\usepackage{color,graphicx,ulem,bm}

\newcommand{\mbf}[1]{\mathbf{#1}}

\newcommand{\pd}[2]{\frac{\partial #1}{\partial #2}}
\newcommand{\unit}[1]{\mathbf{\hat{#1}}}

\newcommand{\vth}[1][s]{\ensuremath{v_{\mathrm{th}_#1}}}

\renewcommand{\eqref}[1]{Eq.\ (\ref{#1})}

\renewcommand{\vth}{\ensuremath{ v_{\mathrm{th}} } }

\newcommand{\vvol}{d^{3}\mbf{v}}

\newcommand{\vpa}{v_{\parallel}}
\newcommand{\vpe}{v_{\perp}}

\newcommand{\beq}{\begin{equation}}
\newcommand{\eeq}{\end{equation}}
\newcommand{\grad}{\nabla}

\newcommand{\rhoi}{\rho_i}

\newcommand{\lpar}{\ell_{\parallel}}
\newcommand{\lperp}{\ell_{\perp}}

\newcommand{\phin}{\Phi}
\newcommand{\lo}{\lperp^o}
\newcommand{\lxo}{\ell_x^o}
\newcommand{\lyo}{\ell_y^o}
\newcommand{\lpo}{\ell_{\parallel}^o}
\newcommand{\phio}{\Phi_o}

\begin{document}

\title{Critically balanced ion temperature gradient turbulence in fusion plasmas}

\author{M.\ Barnes}
\email{michael.barnes@physics.ox.ac.uk}
\affiliation{Rudolf Peierls Centre for Theoretical Physics, University of Oxford, Oxford OX1 3NP, UK}
\affiliation{Euratom/CCFE Fusion Association, Culham Science Centre, Abingdon OX14 3DB, UK}
\author{F.\ I.\ Parra}
\affiliation{Rudolf Peierls Centre for Theoretical Physics, University of Oxford, Oxford OX1 3NP, UK}
\author{A.\ A.\ Schekochihin}
\affiliation{Rudolf Peierls Centre for Theoretical Physics, University of Oxford, Oxford OX1 3NP, UK}

\begin{abstract}

Scaling laws for ion temperature gradient driven turbulence in magnetized toroidal plasmas
are derived and compared with direct numerical simulations.  Predicted dependences of 
turbulence fluctuation amplitudes, spatial scales, and resulting heat fluxes on temperature gradient and magnetic field line pitch are found
to agree with numerical results in both the driving and inertial ranges.  Evidence is provided to support
the critical balance conjecture that parallel streaming and nonlinear perpendicular decorrelation times are comparable
at all spatial scales, leading to a scaling relationship between parallel and perpendicular spatial scales.  
This indicates that even strongly magnetized plasma turbulence is intrinsically three-dimensional.

\end{abstract}



\maketitle

\paragraph{Introduction.}
Microscale turbulence is a ubiquitous feature of the plasmas used for magnetic confinement fusion.
It is driven by kinetic instabilities feeding predominantly off a strong mean gradient in the ion temperature, 
and it is responsible for the majority of particle and heat transport observed in experiment.  As with neutral fluid 
and magnetohydrodynamic turbulence, 
exact analytical results for kinetic plasma turbulence are rare, and numerical simulations are costly.  
Phenomenological scaling laws are thus useful for guiding simulation and providing
gross predictions of plasma behavior in a multi-dimensional parameter space.

Experimental, numerical, and analytical results (cf. ~\cite{pettyPoP04,kotschPoP95,kinseyPoP06,connorPPCF94}) have long 
been used to predict the dependence of turbulent fluxes on the mean plasma
gradients and on the magnetic field configuration.  However, scalings based on empirical observations provide limited physical insight,
and the theoretical predictions, which are predominantly based on linear or quasilinear arguments, are not sufficiently detailed to be easily
falsifiable.  A more detailed examination of the properties of kinetic plasma turbulence has been conducted for scales smaller than the ion
Larmor radius~\cite{schekPPCF08,tatsunoPRL09,plunkJFM10}, but it is the ion temperature gradient (ITG) driven turbulence above the 
Larmor scale that is most relevant for heat transport in fusion devices (cf.~\cite{ottavianiPR97}).
Recent advances in plasma fluctuation measurements~\cite{whitePoP08,hennequinNF06} have provided turbulence spectra in this
scale range; direct numerical simulations have also calculated spectra~\cite{goerlerPoP08} and studied energy injection, transfer, and 
dissipation~\cite{banonPRL11,hatchPRL11}.  

In this Letter, we propose a phenomenological scaling theory of ITG turbulence.  A number of simple, physically-motivated conjectures about the nature of 
this turbulence are formulated and applied to obtain fluctuation spectra from the driving scale to the ion Larmor scale.  We then derive predictions for the dependence 
of heat flux on plasma current and ion temperature gradient.  Numerical results are presented to support our predictions and justify our conjectures.

%Numerical~\cite{kotschPoP95,kinseyPoP06} and experimental~\cite{luceNF02,pettyPoP04} results have long been used to obtain empirical 
%scalings of plasma turbulence quantities with parameters such as plasma current, magnetic shear, collisionality, and mean density 
%and temperature gradients.  Reduced theoretical models have also been employed to obtain analytical scalings for gross plasma
%quantities such as thermal diffusivities and confinement time~\cite{connorPPCF94}.  In this Letter we use a first-principles, gyrokinetic model
%to obtain analytical scaling laws for ion temperature gradient (ITG) driven turbulence.  Predictions are made for the dependence on safety factor and ion temperature
%gradient of fluctuation amplitudes, characteristic turbulence spatial scales, and ion heat flux.  Additionally, a scaling for the turbulence fluctuation 
%spectrum is given, and the critical balance~\cite{goldreichApJ95,nazarenkoJFM11} conjecture is used to relate parallel and perpendicular length scales.  
%The analytical predictions are consistent with experimental fluctuation measurements~\cite{hennequinNF06} and direct numerical simulations
%presented in this Letter.

\paragraph{Gyrokinetic turbulence.}

Plasma fluctuations in a strong mean magnetic field are anisotropic with respect to the mean field direction and
have typical frequencies that are small compared to the ion Larmor frequency, $\Omega_i$. 
Such fluctuations are correctly described by the gyrokinetic approximation~\cite{cattoPoF78}.
It assumes $\tau\Omega_i \gg 1$, $\lperp/\lpar \sim \rhoi/L \ll 1$, and $\rhoi/\lperp\sim 1$, where $\tau$ is the fluctuation time,
$\rhoi$ the ion Larmor radius, $L$ the characteristic length scale of the mean dynamics, and $\lpar$ and $\lperp$ are the fluctuation length scales
parallel and perpendicular to the mean field, respectively.  Averaging over the fast Larmor gyration eliminates gyroangle dependence.
For the remaining phase space coordinates, we choose $(\mbf{R},E,\mu)$, where $\mbf{R}$ is 
the position of the center of a particle's Larmor orbit, $E=mv^2/2$ is the kinetic energy of a particle, and $\mu=mv_{\perp}^2/2B$ is the magnetic moment, 
with $m$ particle mass, $v$ particle speed, $v_{\perp}$ its perpendicular component, and $B$ mean magnetic field strength.  

We restrict our attention to ion scale turbulence with $\rhoi/\lperp < 1$, for which the guiding center position, $\mbf{R}$, and actual
particle position, $\mbf{r}$, are approximately equal.  The equation describing the evolution of electrostatic 
fluctuations in the absence of sonic flows is
\beq
\begin{split}
\pd{}{t}\Big(h_s&-\frac{Z_s e\varphi}{T_s}F_{M,s}\Big)+\left(\mbf{\vpa}+\mbf{v}_{M,s}\right)\cdot\grad h_s \\
&+ \mbf{v}_{E}\cdot\grad \left(F_{M,s}+h_s\right)=C[h_s],
\end{split}
\label{eqn:gk}
\eeq
where $\delta f_s=h_s-(Z_s e\varphi/T_s)F_{M,s}$ describes the distribution of particle positions and velocities for species $s$, $t$ is time, 
$\mbf{v}_{M,s}=(\unit{b}/\Omega_s)\times(\vpa^2 \unit{b}\cdot\grad\unit{b} + (\vpe^2/2)\grad B/B)$ is the magnetic drift velocity,
$\mbf{v}_{E}=(c/B)\unit{b}\times\grad \varphi$ is the $\mbf{E}\times\mbf{B}$ drift velocity, $\unit{b}$ is the unit vector along the mean field,
$Z_s e$ is the species charge, $e$ is the proton charge, 
$T_s$ is the temperature, $F_{M,s}$ is a Maxwellian distribution of velocities, and $C$ is a Fokker-Planck collision operator.  

The electrostatic potential $\varphi$ is obtained by imposing quasineutrality, $\sum_s Z_s e n_s = 0$, which can be written
\beq
\sum_s Z_s \int\vvol \left(h_s - \frac{Z_s e\varphi}{T_s}F_{M,s}\right) = 0,
\label{eqn:qn}
\eeq
where $n_s$ is species density.
Assuming $\int \vvol h_i\sim \vth^3 h_i$, where $i$ denotes the main ion species and $\vth$ its thermal
speed, Eq.~(\ref{eqn:qn}) gives $h_i/F_{M,i} \sim Z_i e \varphi/T_i$.

\paragraph{Critical Balance.}

Because the turbulence we are considering is anisotropic, dimensional analysis alone is not sufficient
to determine scalings of the fluctuation amplitudes with both $\lpar$ and $\lperp$.  To fix the ratio of $\lpar$ to $\lperp$, we make a conjecture known as 
critical balance~\cite{goldreichApJ95}:  \textit{The characteristic time associated with particle streaming and wave propagation along 
the mean field, $\lpar/\vth$, is comparable to the nonlinear decorrelation time at each scale.}
This is motivated by the causality constraint:  two points along the mean field can be correlated only if
information can propagate between these points in the time it takes turbulence to decorrelate in the
plane perpendicular to the field.  It gives us a relation between parallel and perpendicular spatial scales:
\beq
\frac{\vth}{\lpar} \sim \tau_{nl}^{-1} \sim \frac{\vth}{R}\frac{\rhoi^2}{\ell_x\ell_y}\phin_{\ell},
\label{eqn:cb}
\eeq
where $\tau_{nl}$ is the nonlinear decorrelation time, $R$ is the major radius of the torus,
$\phin_{\ell}\equiv (e\varphi_{\ell}/T)(R/\rhoi)$ and 
$\varphi_{\ell}\equiv \varphi(\mbf{r}+\bm{\ell})-\varphi(\mbf{r})$.
The subscripts $x$ and $y$ refer to the coordinates in the plane perpendicular to the mean field, with $x$ labeling
surfaces of constant magnetic flux and $y$ labeling field lines within a constant flux surface.

\paragraph{Outer Scale.}

We define the outer scale as the scale for which the time associated with the linear drive is comparable to the nonlinear
decorrelation time:
\beq
\tau_{nl}^{-1} \sim \omega_*^o \sim \frac{\rhoi \vth}{\ell_y^o L_T},
\label{eqn:omega*}
\eeq
where $\omega_*$ is the frequency associated with the ITG drive, $L_T$ is the ITG scale length, and
$o$ labels outer scale quantities.  The outer scale corresponds to the injection range, which contains the turbulence
amplitude peak.  We conjecture that \textit{the characteristic parallel length at the outer scale, $\lpo$, is the parallel system size}.
In toroidal plasmas, this is the distance along the mean field from the outside to the inside of the
torus.  This distance, known as the connection length, is $qR$, where the safety factor $q$ measures the pitch
of the field lines.  Thus,
\beq
\lpo \sim qR.
\label{eqn:lparqr}
\eeq
Combining relations (\ref{eqn:cb})-(\ref{eqn:lparqr}) provides a prediction
for the dependence of $\lyo$ on $q$ and $R/L_T$~\footnote{Scaling~(\ref{eqn:Ly}) can also be obtained from linear
analysis of the slab ITG mode, as shown in~\cite{ottavianiPR97}.}
%However, \cite{ottavianiPR97} proceeds by making the contradictory
%assumption that the curvature-driven ITG mode controls the nonlinear state.  This leads to inconsistent transport scalings (that differ from
%the ones we later derive).}
\beq
\frac{\lyo}{\rhoi} \sim \frac{qR}{L_T}.
\label{eqn:Ly}
\eeq
Since $qR/L_T\gg 1$ in fusion plasmas, the ratio $\rhoi/\lyo \ll 1$.  Thus,
there is a range of scales between the outer scale and the Larmor scale, below which
kinetic damping effects are expected to become significant.

The final piece of information necessary to determine scalings for the fluctuation
amplitude at the outer scale, $\phio$, is
a relationship between $\lxo$ and $\lyo$.  We conjecture that
\textit{the length scales in the perpendicular plane, $\ell_x$ and $\ell_y$, are comparable
at all scales}:  $\ell_x\sim\ell_y\sim \lperp$.
Indeed, one might argue that $\ell_x$ is set nonlinearly
at the outer scale through the shearing of radially extended eddies by zonal flow: 
$\ell_x^{-1}\sim\ell_y^{-1}(S_{\textnormal{ZF}}\tau_{nl})$, where $S_{\textnormal{ZF}}$ is the
shearing rate at the outer scale due to zonal flow.  For strong ITG turbulence, one expects
$S_{\textnormal{ZF}}\tau_{nl}\sim 1$~\cite{cowleyPoFB91}, so that $\ell_x\sim\ell_y$ is satisfied.

Taking $\ell_x\sim\ell_y$ and using (\ref{eqn:cb}) and (\ref{eqn:Ly}) in (\ref{eqn:omega*}) gives
\beq
\phio \sim \frac{\lxo}{\rhoi}\frac{R}{L_T} \sim q \left(\frac{R}{L_T}\right)^2.
\label{eqn:phio}
\eeq
If we assume that the ion turbulent heat flux through volume $V$, $Q_i\equiv V^{-1}\int d^3\mbf{r} \int d^3\mbf{v} \left(\mbf{v}_E\cdot\grad x\right)\left(m_i v^2/2\right)\delta f_i$,
is dominated by the contribution from the outer scale, relations
(\ref{eqn:Ly}) and (\ref{eqn:phio}) imply a scaling for $Q_i$:
\beq
\frac{Q_i}{n_iT_i\vth}\left(\frac{R}{\rhoi}\right)^2 \equiv \tilde{Q}_i \sim \frac{\rhoi}{\lyo}  \phio^2 \sim q\left(\frac{R}{L_T}\right)^3.
\label{eqn:qflx}
\eeq
The $R/L_T$ scaling is only valid for sufficiently large $R/L_T$ because our simple analysis ignores the finite critical temperature
gradient associated with the ITG instability.
%The predicted scaling of turbulence amplitude, and thus heat flux, with $R/L_T$ is only valid for sufficiently large $R/L_T$
%because our simple analysis does not account for the existence of the critical temperature gradient associated with ITG turbulence.

\begin{figure}
\includegraphics[height=2.1in]{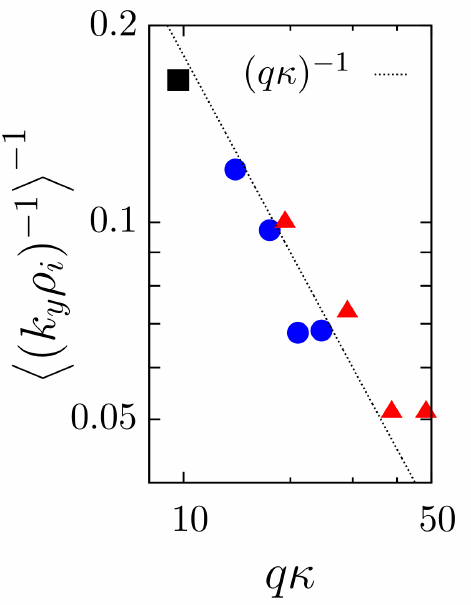}
\includegraphics[height=2.1in]{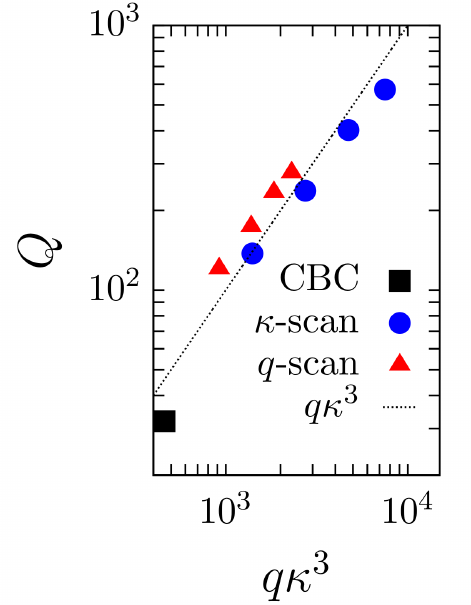}
\caption{(a) Expectation value of $k_{y}\rhoi$ versus $q\kappa$, where $\kappa\equiv R/L_T$. (b) Normalized heat flux versus $q\kappa^3$. 
Lines show the predicted scalings~(\ref{eqn:Ly}) and (\ref{eqn:qflx}).}
\label{fig:qflx}
\end{figure}

\paragraph{Inertial Range.}

We now consider the range of scales between $\lo$ and $\rhoi$ and conjecture that it is an inertial range:
\textit{There is no significant dissipation or driving at scales between $\lo$ and $\rhoi$}.
%\begin{list}{\textit{Conjecture 5.}}
%{\setlength{\rightmargin}{\leftmargin}}
%\item There is no significant dissipation or driving at scales between $\rhoi$ and $\lo$.
%\end{list}
This conjecture is to be checked \textit{a posteriori}.  To determine the spectrum in the inertial range, 
we identify the free energy $W=V^{-1}\sum_s\int d^3\mbf{r}\int d^3\mbf{v} T_s\delta f_s^2/F_{M,s}$
as a nonlinear invariant~\cite{schekPPCF08} and consider scale-by-scale energy balance.  Because we are in an inertial
range, the flux of free energy, $W_{\ell}/\tau_{nl}$, must be independent of $\lperp$:
\beq
\frac{1}{n_iT_i}\frac{W_{\ell}}{\tau_{nl}}\sim\left(\frac{\rhoi}{R}\right)^2\frac{\vth}{R}\frac{\rhoi^2}{\ell_x\ell_y}\Phi_{\ell}^3\sim\textnormal{constant}.
\label{eqn:W}
\eeq
The dependence of $\Phi_{\ell}$ on $q$ and $R/L_T$ is obtained by solving relation (\ref{eqn:W}) for $\Phi_{\ell}$,
matching to $\phio$ [see (\ref{eqn:phio})], using relation (\ref{eqn:Ly}), and assuming isotropy:
\beq
\Phi_{\ell} \sim \phio \left(\frac{\lperp}{\lyo}\right)^{2/3}\sim q^{1/3}\left(\frac{R}{L_T}\right)^{4/3}\left(\frac{\lperp}{\rhoi}\right)^{2/3}.
\label{eqn:phil}
\eeq
Parseval's theorem relates $\Phi_{\ell}$ to the Fourier coefficient $\Phi_{\bm{k}}$, giving a scaling
for the 1D fluctuation spectrum, $E(k_{y})$, defined so that $\int dk_y \ \rhoi E(k_y)=V^{-1}\int d^3\mbf{r} \ \Phi^2$:
\beq
E(k_y) \sim k_y\rhoi\left|\Phi_{\bm{k}}\right|^2 \sim q^{2/3}\left(\frac{R}{L_T}\right)^{8/3}\left(k_{\perp}\rhoi\right)^{-7/3},
\label{eqn:spectrum}
\eeq
where $k_{\perp}=2\pi/\lperp$.  Using (\ref{eqn:phil}) to evaluate $\tau_{nl}$, we find $\omega_* \sim (\lperp/\lo)^{1/3}\tau_{nl}^{-1}$,
confirming that the drive is subdominant to the energy transfer in the inertial range~\footnote{It is shown in~\cite{hatchPRL11} that 
there can be significant dissipation above the outer scale, but that does not affect the inertial range arguments given here.}.

Using relation (\ref{eqn:phil}) and applying critical balance (\ref{eqn:cb}) gives the scaling of $\lpar$ with $\lperp$:
\beq
\frac{\lpar}{qR} \sim \left(\frac{\lperp}{\rhoi}\frac{L_T}{qR}\right)^{4/3}.
\label{eqn:lpar}
\eeq
Converting this into a scaling for $\phin_{\ell}$, we obtain the parallel structure function
$\phin_{\ell}^2\sim q(R/L_T)^4(\lpar/R)$.

\begin{figure}
\includegraphics[height=2.1in]{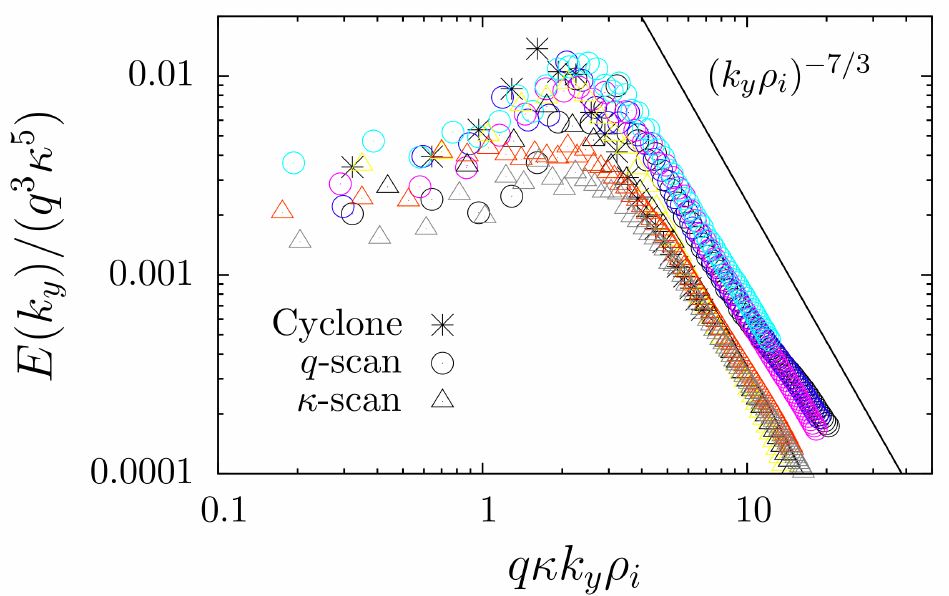}
\caption{Electrostatic fluctuation spectra, $E(k_y)$, normalized to the predicted scaling~(\ref{eqn:spectrum}) at the outer scale.
We have normalized $k_{y}\rhoi$ using the outer scale relationship (\ref{eqn:Ly}).  
Asterisks denote the Cyclone Base Case ($q=1.4$, $\kappa\equiv R/L_T=6.9$); circles and triangles denote simulations where $q$ and $\kappa$ vary from $2.8$ to $7.0$ and $10.0$ to $17.5$, respectively.  The solid line gives the predicted inertial range scaling (\ref{eqn:spectrum}).}
\label{fig:kphi2}
\end{figure}

\paragraph{Numerical results.}

We next compare our scaling relations with numerical results obtained using
the gyrokinetic code \texttt{GS2}~\cite{dorlandPRL00}.
We restrict our attention to electrostatic fluctuations with perturbed electron density 
$\delta n_e/n_e= e(\varphi-\overline{\varphi})/T_e$, 
with the overline denoting a flux surface average.  The magnetic 
geometry is as in the widely-benchmarked
Cyclone Base Case (CBC)~\cite{dimitsPoP00}: unshifted, circular magnetic-flux surface, 
with $r/R=0.18$, $\hat{s}=d\ln q/d \ln r=0.8$, and $R/L_n=2.2$,
where $r$ is the minor radius of the flux surface, and $L_n$ is the density gradient
scale length.  The parameters $q$ and $\kappa\equiv R/L_T$ were varied over several
simulations to obtain numerical scalings.  These simulations employed a small amount of
upwinding along the magnetic field and hyper-dissipation in $k_{\perp}$~\cite{belli},
cutting off the fluctuation spectra at $k_{\perp}\rhoi \simeq 1$.

In Fig.~\ref{fig:qflx}, we show the $q$ and $\kappa$ dependences of the normalized ion 
heat flux, $\tilde{Q}_i$, and of
\beq
\frac{\lyo}{\rhoi}\sim\left<\left(k_{y}\rhoi\right)^{-1} \right> = \frac{\sum_{k_x,k_y} \left(k_{y}\rhoi\right)^{-1}\left|\Phi(k_x,k_y)\right|^2}{\sum_{k_x,k_y} \left|\Phi(k_x,k_y)\right|^2},
\eeq
which is a good measure of the outer scale, provided the spectrum is sufficiently steep.  
The simulations agree remarkably well with our predicted scalings (\ref{eqn:Ly}) and 
(\ref{eqn:qflx})~\footnote{The heat fluxes shown in Fig.~\ref{fig:qflx} are significantly larger than those 
reported in~\cite{dimitsPoP00} at large $\kappa$.  In order to obtain our results, it was necessary
to use much larger simulation domains ($\sim300\pi\rhoi$) than those considered in~\cite{dimitsPoP00}.}.

The one-dimensional spectrum, $E(k_y)$, is plotted vs.~$k_y\rhoi$ in
Fig.~\ref{fig:kphi2} for several $q$ and $\kappa$ values.  At scales smaller than the outer scale,
all spectra follow the same power law, which agrees with (\ref{eqn:spectrum}).

\begin{figure}
\includegraphics[height=2.2in]{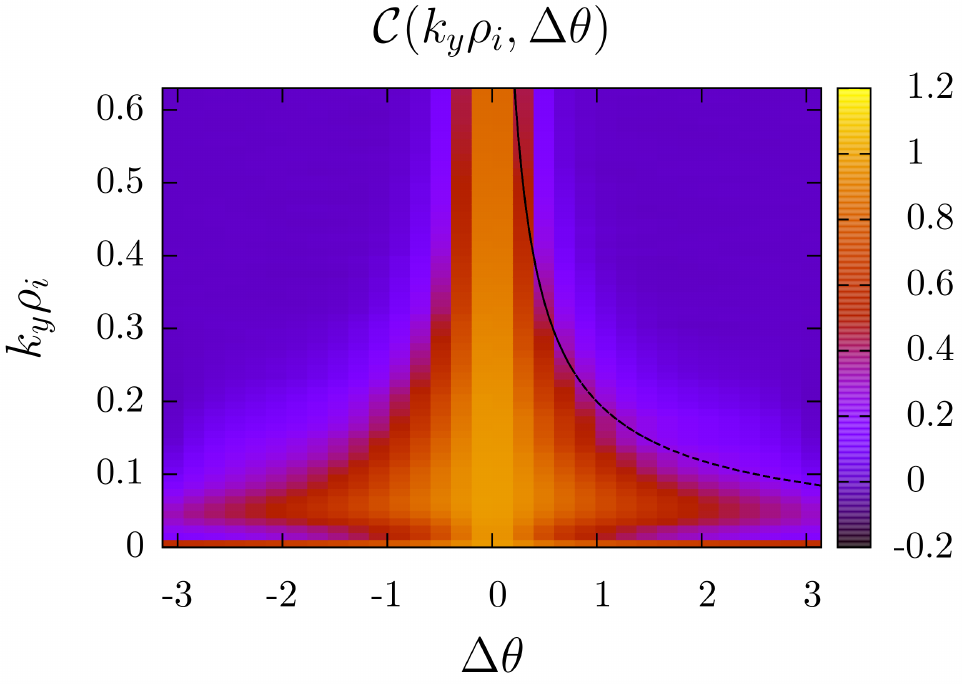}
\caption{Correlation function, $\mathcal{C}$ (\ref{eqn:cfnc}).  The solid black line
is $\Delta\theta\propto k_{y}^{-4/3}$, the critical balance scaling (\ref{eqn:lpar}).}
\label{fig:cfnc}
\end{figure}

\begin{figure}
\includegraphics[height=2.1in]{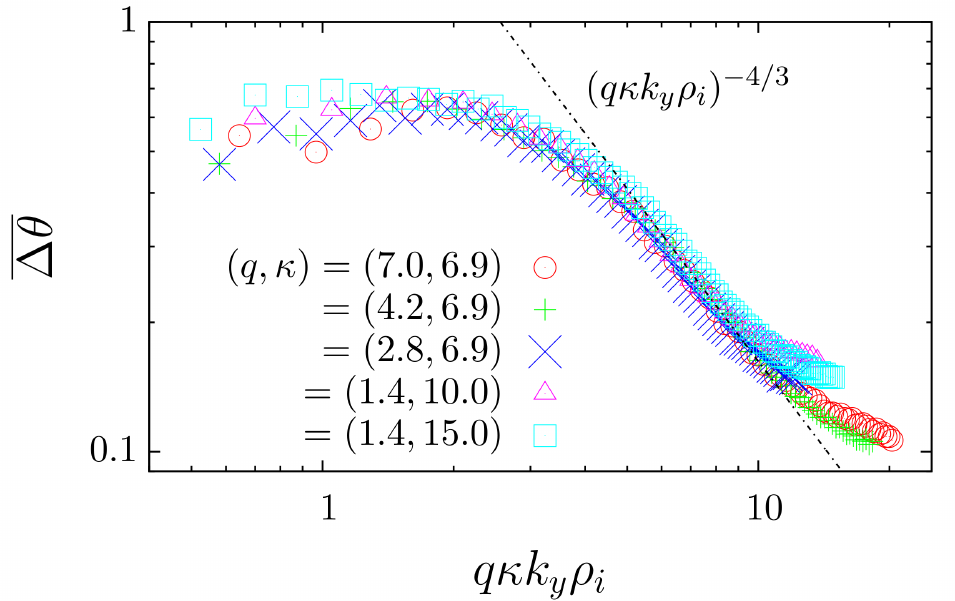}
\caption{Scaling of poloidal correlation length with $q\kappa k_y\rhoi$ for several simulations with varying $q$ and $\kappa\equiv R/L_T$.  
The dashed line indicates the critical balance scaling~(\ref{eqn:lpar}).}
\label{fig:clength}
\end{figure}

Our predictions for the critical balance scaling~(\ref{eqn:lpar}) are tested in Figs.~\ref{fig:cfnc}
and~\ref{fig:clength}.  The parallel correlation function,
\beq
\mathcal{C}(k_y,\Delta\theta) \equiv \frac{\sum_{k_x} \Phi(k_x,k_y,\theta=0)\Phi^*(k_x,k_y,\theta=\Delta\theta)}
{\sum_{k_x} \left|\Phi(k_x,k_y,\theta=0)\right|^2},
\label{eqn:cfnc}
\eeq
is plotted in Fig.~\ref{fig:cfnc} for the simulation with $q=4.2$ and $\kappa=6.9$.  Here $\theta$ is the poloidal angle of the torus
so that $\Delta \theta \sim \lpar/qR$, with $\theta=0$ at the outermost point on the flux surface.  Our prediction (\ref{eqn:lpar})
is given by the black line, which fits the data.  The $qR$-normalized correlation length, 
$\overline{\Delta\theta}(k_y)\equiv \int d(\Delta \theta) \mathcal{C}(k_y,\Delta\theta)$,
is plotted in Fig.~\ref{fig:clength} for multiple $q$ and $\kappa$ values.  At sufficiently large $k_y\rhoi$, the data
follows the power law (\ref{eqn:lpar}).  Note that $\overline{\Delta \theta}(k_y)$ peaks at a value of approximately unity,
in agreement with (\ref{eqn:lparqr}).

%\paragraph{Caveats.}

%We have numerically tested most of our conjectures, but we have yet to justify (\ref{eqn:isotropy}) and (\ref{eqn:W}).
%Relation (\ref{eqn:isotropy}) is a statement of isotropy of turbulence in the inertial range; at the outer scale
%we note that the linear instability injects generalized energy at long wavelength in $x$, and we argue
%that $\ell_x$ is set nonlinearly by a combination of magnetic shear and zonal flow shear.  Thus
%\beq
%\ell_x^{-1} \sim \ell_y^{-1}\left(\frac{\tau_{nl}}{\tau_{zf}} + \hat{s}\theta\right),
%\eeq
%where $\tau_{zf}^{-1}$ is the shearing rate due to zonal flow.  For $\theta$ and $\hat{s}$ order unity
%and $\tau_{nl}\sim \tau_{zf}$, we have $\ell_x\sim\ell_y$.

\paragraph{Sub-Larmor scales.}

\begin{figure}
\includegraphics[height=2.2in]{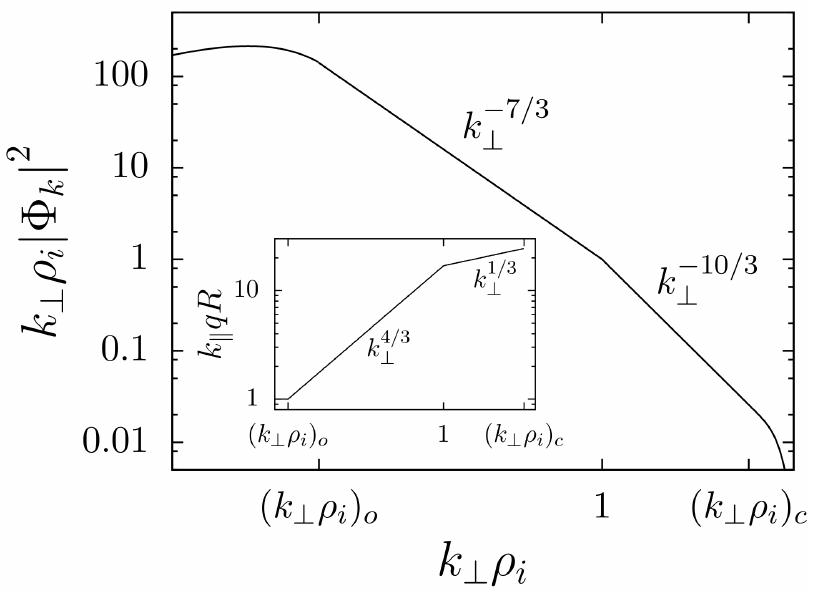}
\caption{Cartoon of the fluctuation spectrum from the outer scale, $(k_{\perp}\rhoi)_o$ [Eq. (\ref{eqn:Ly})],
to the dissipation scale, $(k_{\perp}\rhoi)_c$ [Eq. (\ref{eqn:kcut})].  Scalings for $k_{\perp}\rhoi>1$
are taken from~\cite{schekPPCF08}.}
\label{fig:cascade}
\end{figure}

Using arguments similar to those presented above, scalings of $E(k_\perp)\sim k_{\perp}^{-10/3}$ and 
$(k_{\perp}\rhoi)_c \sim \textnormal{Do}^{3/5}$ were obtained for 
the sub-Larmor scales in~\cite{schekPPCF08} and verified numerically in~\cite{tatsunoPRL09}.  Here the subscript $c$ denotes 
the cutoff wavenumber, and the Dorland number $\textnormal{Do}\equiv (\tau_{\rhoi} \nu_{ii})^{-1}$~\cite{schekPPCF08,tatsunoPRL09,plunkJFM10} 
is the kinetic plasma turbulence analog of the Reynolds number, 
with $\tau_{\rhoi}$ the nonlinear time at $\rhoi$ and $\nu_{ii}$ the ion-ion collision frequency.  Using relations (\ref{eqn:phil}) and (\ref{eqn:cb}),
we find:
\beq
\left(k_{\perp}\rhoi\right)_c \sim \textnormal{Do}^{3/5}\sim q^{1/5}\left(\frac{R}{L_T}\right)^{4/5}\left(\frac{\vth}{\nu_{ii} R}\right)^{3/5}.
\label{eqn:kcut}
\eeq
Combining the results of~\cite{schekPPCF08} with those given here provides a complete picture of the spectra of ITG turbulence 
from the driving to dissipation scales, shown in Fig.~\ref{fig:cascade}.

\paragraph{Discussion.}

The main results obtained in this Letter are: a scaling of heat flux (\ref{eqn:qflx}) and dissipation
scale (\ref{eqn:kcut}) with $q$ and $R/L_T$;
a power law scaling for the electrostatic fluctuation spectrum (\ref{eqn:spectrum}); and a relationship between
parallel and perpendicular length scales (\ref{eqn:lpar}).  The heat flux scaling, confirmed
numerically (see Fig.~\ref{fig:qflx}), has a stronger dependence on $R/L_T$ than is usually
assumed by reduced models for turbulent transport.  While this has little effect for near-marginal plasma
turbulence, it may be significant in the vicinity of steep gradient regions.

The power law prediction for the turbulence spectrum (\ref{eqn:spectrum}), also
confirmed numerically (see Fig.~\ref{fig:kphi2}), appears to be consistent with recent experimental fluctuation 
measurements~\cite{hennequinNF06}.  Along with the predictions for the dissipation scale (\ref{eqn:kcut}),
the spectrum could be used to design large eddy simulations for gyrokinetics~\cite{banonPRL11} and
to guide resolution choices for direct numerical simulations.  The critical balance conjecture (\ref{eqn:cb}) has proven to be robustly satisfied for the ITG turbulence
considered here (see Figs.~\ref{fig:cfnc} and~\ref{fig:clength}).  
This indicates that ITG turbulence is an inherently three-dimensional phenomenon.  

%Poorly resolving parallel dynamics may thus lead to inadequate modeling of the turbulence.

%In conclusion, we have shown that ITG turbulence exhibits many of the universal properties
%(inertial range, critical balance, etc.) %~\cite{kolmogorovCRAS41,goldreichApJ95,nazarenkoJFM11}
%characteristic of turbulent systems and that simple scaling arguments can provide accurate
%predictions for turbulence spectra and transport.

We thank W. Dorland, S. C. Cowley, and G. W. Hammett for useful discussions, which 
were enabled by travel support from the Leverhulme Trust International Network for Magnetized Plasma Turbulence.
M.B. was supported by the Oxford-Culham Fusion Research Fellowship, F.I.P. was supported by EPSRC, and computing time was provided
by HPC-FF (J\"ulich).

%\begin{figure}
%\includegraphics[height=2.2in]{figs/phi2avg_by_mode_nozf.pdf}
%\caption{Spectrum of $\left|\Phi\right|^2$ in the perpendicular wavenumber space.  The solid black line is the line
%for which the two perpendicular wavenumbers are the same.}
%\label{fig:phi2_by_mode}
%\end{figure}

%\bibliography{/Users/michael/Documents/tex_stuff/general}

\begin{thebibliography}{19}
\expandafter\ifx\csname natexlab\endcsname\relax\def\natexlab#1{#1}\fi
\expandafter\ifx\csname bibnamefont\endcsname\relax
  \def\bibnamefont#1{#1}\fi
\expandafter\ifx\csname bibfnamefont\endcsname\relax
  \def\bibfnamefont#1{#1}\fi
\expandafter\ifx\csname citenamefont\endcsname\relax
  \def\citenamefont#1{#1}\fi
\expandafter\ifx\csname url\endcsname\relax
  \def\url#1{\texttt{#1}}\fi
\expandafter\ifx\csname urlprefix\endcsname\relax\def\urlprefix{URL }\fi
\providecommand{\bibinfo}[2]{#2}
\providecommand{\eprint}[2][]{\url{#2}}

\bibitem[{\citenamefont{Petty et~al.}(2004)\citenamefont{Petty, Kinsey, and
  Luce}}]{pettyPoP04}
\bibinfo{author}{\bibfnamefont{C.~C.} \bibnamefont{Petty}},
  \bibinfo{author}{\bibfnamefont{J.~E.} \bibnamefont{Kinsey}},
  \bibnamefont{and} \bibinfo{author}{\bibfnamefont{T.~C.} \bibnamefont{Luce}},
  \bibinfo{journal}{Phys. Plasmas} \textbf{\bibinfo{volume}{11}},
  \bibinfo{pages}{1011} (\bibinfo{year}{2004}).

\bibitem[{\citenamefont{Kotschenreuther
  et~al.}(1995{\natexlab{a}})\citenamefont{Kotschenreuther et~al.}}]{kotschPoP95}
\bibinfo{author}{\bibfnamefont{M.}~\bibnamefont{Kotschenreuther}} \bibnamefont{et~al.},
  \bibinfo{journal}{Phys. Plasmas} \textbf{\bibinfo{volume}{2}},
  \bibinfo{pages}{2381} (\bibinfo{year}{1995}{\natexlab{a}}).

\bibitem[{\citenamefont{Kinsey et~al.}(2006)\citenamefont{Kinsey, Waltz, and
  Candy}}]{kinseyPoP06}
\bibinfo{author}{\bibfnamefont{J.~E.} \bibnamefont{Kinsey}},
  \bibinfo{author}{\bibfnamefont{R.~E.} \bibnamefont{Waltz}}, \bibnamefont{and}
  \bibinfo{author}{\bibfnamefont{J.}~\bibnamefont{Candy}},
  \bibinfo{journal}{Phys. Plasmas} \textbf{\bibinfo{volume}{13}},
  \bibinfo{pages}{022305} (\bibinfo{year}{2006}).

\bibitem[{\citenamefont{Connor and Wilson}(1994)}]{connorPPCF94}
\bibinfo{author}{\bibfnamefont{J.~W.} \bibnamefont{Connor}} \bibnamefont{and}
  \bibinfo{author}{\bibfnamefont{H.~R.} \bibnamefont{Wilson}},
  \bibinfo{journal}{Plasma Phys. Control. Fusion}
  \textbf{\bibinfo{volume}{36}}, \bibinfo{pages}{719} (\bibinfo{year}{1994}).

\bibitem[{\citenamefont{Schekochihin et~al.}(2008)\citenamefont{Schekochihin et~al.}}]{schekPPCF08}
\bibinfo{author}{\bibfnamefont{A.~A.} \bibnamefont{Schekochihin}} \bibnamefont{et~al.},
  \bibinfo{journal}{Plasma Phys. Control. Fusion}
  \textbf{\bibinfo{volume}{50}}, \bibinfo{pages}{124024}
  (\bibinfo{year}{2008});
\bibinfo{author}{\bibfnamefont{A.~A.} \bibnamefont{Schekochihin}} \bibnamefont{et~al.},
  \bibinfo{journal}{Astrophys. J. Suppl.} \textbf{\bibinfo{volume}{182}},
  \bibinfo{pages}{310} (\bibinfo{year}{2009}).

\bibitem[{\citenamefont{Tatsuno et~al.}(2009)\citenamefont{Tatsuno et~al.}}]{tatsunoPRL09}
\bibinfo{author}{\bibfnamefont{T.}~\bibnamefont{Tatsuno}} \bibnamefont{et~al.},
  \bibinfo{journal}{Phys. Rev. Lett.} \textbf{\bibinfo{volume}{103}},
  \bibinfo{pages}{015003} (\bibinfo{year}{2009}).

\bibitem[{\citenamefont{Plunk et~al.}(2010)\citenamefont{Plunk et~al.}}]{plunkJFM10}
\bibinfo{author}{\bibfnamefont{G.~G.} \bibnamefont{Plunk}} \bibnamefont{et~al.},
  \bibinfo{journal}{J. Fluid Mech.} \textbf{\bibinfo{volume}{664}},
  \bibinfo{pages}{407} (\bibinfo{year}{2010}).

\bibitem[{\citenamefont{Ottaviani et~al.}(1997)\citenamefont{Ottaviani et~al.}}]{ottavianiPR97}
\bibinfo{author}{\bibfnamefont{M.}~\bibnamefont{Ottaviani}} \bibnamefont{et~al.},
  \bibinfo{journal}{Phys. Rep.} \textbf{\bibinfo{volume}{283}},
  \bibinfo{pages}{121} (\bibinfo{year}{1997}).

\bibitem[{\citenamefont{White et~al.}(2008)\citenamefont{White et~al.}}]{whitePoP08}
\bibinfo{author}{\bibfnamefont{A.~E.} \bibnamefont{White}} \bibnamefont{et~al.},
\bibinfo{journal}{Phys. Plasmas}
  \textbf{\bibinfo{volume}{15}}, \bibinfo{pages}{056116}
  (\bibinfo{year}{2008}).

\bibitem[{\citenamefont{Hennequin et~al.}(2006)\citenamefont{Hennequin et~al.}}]{hennequinNF06}
\bibinfo{author}{\bibfnamefont{P.}~\bibnamefont{Hennequin}} \bibnamefont{et~al.},
  \bibinfo{journal}{Nucl. Fusion} \textbf{\bibinfo{volume}{46}},
  \bibinfo{pages}{S771} (\bibinfo{year}{2006}).

\bibitem[{\citenamefont{G\"orler and Jenko}(2008)}]{goerlerPoP08}
\bibinfo{author}{\bibfnamefont{T.}~\bibnamefont{G\"orler}} \bibnamefont{and}
  \bibinfo{author}{\bibfnamefont{F.}~\bibnamefont{Jenko}},
  \bibinfo{journal}{Phys. Plasmas} \textbf{\bibinfo{volume}{15}},
  \bibinfo{pages}{102508} (\bibinfo{year}{2008});
\bibinfo{author}{\bibfnamefont{A.}~\bibnamefont{Casati}} \bibnamefont{et~al.},
\bibinfo{journal}{Phys. Rev. Lett.}
  \textbf{\bibinfo{volume}{102}}, \bibinfo{pages}{165005}
  (\bibinfo{year}{2009}).
  
\bibitem[{\citenamefont{{n}on Navarro et~al.}(2011)\citenamefont{Ba\~non Navarro et~al.}}]{banonPRL11}
\bibinfo{author}{\bibfnamefont{A.}~\bibnamefont{Ba\~non Navarro}} \bibnamefont{et~al.},
  \bibinfo{journal}{Phys. Rev. Lett.} \textbf{\bibinfo{volume}{106}},
  \bibinfo{pages}{055001} (\bibinfo{year}{2011}).

\bibitem[{\citenamefont{Hatch et~al.}(2011)\citenamefont{Hatch et~al.}}]{hatchPRL11}
\bibinfo{author}{\bibfnamefont{D.~R.} \bibnamefont{Hatch}} \bibnamefont{et~al.},
  \bibinfo{journal}{Phys. Rev. Lett.} \textbf{\bibinfo{volume}{106}},
  \bibinfo{pages}{115003} (\bibinfo{year}{2011}).

\bibitem[{\citenamefont{Catto and Tsang}(1978)}]{cattoPoF78}
\bibinfo{author}{\bibfnamefont{P.~J.} \bibnamefont{Catto}},
  \bibinfo{journal}{Plasma Phys.} \textbf{\bibinfo{volume}{20}},
  \bibinfo{pages}{719} (\bibinfo{year}{1978}).

\bibitem[{\citenamefont{Goldreich and Sridhar}(1995)}]{goldreichApJ95}
\bibinfo{author}{\bibfnamefont{P.}~\bibnamefont{Goldreich}} \bibnamefont{and}
  \bibinfo{author}{\bibfnamefont{S.}~\bibnamefont{Sridhar}},
  \bibinfo{journal}{Astrophys. J.} \textbf{\bibinfo{volume}{438}},
  \bibinfo{pages}{763} (\bibinfo{year}{1995});
\bibinfo{author}{\bibfnamefont{S.~V.} \bibnamefont{Nazarenko}}
  \bibnamefont{and} \bibinfo{author}{\bibfnamefont{A.~A.}
  \bibnamefont{Schekochihin}}, \bibinfo{journal}{J. Fluid Mech.}
\textbf{\bibinfo{volume}{677}}, \bibinfo{pages}{134}
  (\bibinfo{year}{2011}).

\bibitem[{\citenamefont{Cowley et~al.}(1991)\citenamefont{Cowley et~al.}}]{cowleyPoFB91}
\bibinfo{author}{\bibfnamefont{S.~C.} \bibnamefont{Cowley}} \bibnamefont{et~al.},
  \bibinfo{journal}{Phys. Fluids B} \textbf{\bibinfo{volume}{3}},
  \bibinfo{pages}{2767} (\bibinfo{year}{1991}).

\bibitem[{\citenamefont{Dorland et~al.}(2000)\citenamefont{Dorland et~al.}}]{dorlandPRL00}
\bibinfo{author}{\bibfnamefont{W.}~\bibnamefont{Dorland}} \bibnamefont{et~al.},
  \bibinfo{journal}{Phys. Rev. Lett}
  \textbf{\bibinfo{volume}{85}}, \bibinfo{pages}{5579} (\bibinfo{year}{2000}).

\bibitem[{\citenamefont{Dimits et~al.}(2000)\citenamefont{Dimits et~al.}}]{dimitsPoP00}
\bibinfo{author}{\bibfnamefont{A.~M.} \bibnamefont{Dimits}} \bibnamefont{et~al.},
\bibinfo{journal}{Phys. Plasmas}
  \textbf{\bibinfo{volume}{7}}, \bibinfo{pages}{969} (\bibinfo{year}{2000}).

\bibitem[{\citenamefont{Belli}(2006)}]{belli}
\bibinfo{author}{\bibfnamefont{E.~A.} \bibnamefont{Belli}}, Ph.D. thesis,
  \bibinfo{school}{Princeton University} (\bibinfo{year}{2006}).

\end{thebibliography}

\end{document}